\documentclass[]{aa}
\def\ueber#1#2{{\setbox0=\hbox{$#1$}%
  \setbox1=\hbox to\wd0{\hss$ #2$\hss}%
  \offinterlineskip
  \vbox{\box1\box0}}{}}

\def\lesssim{\,\lower 1mm \hbox{\ueber{\sim}{<}}\,}
\def\grsim{\,\lower 1mm \hbox{\ueber{\sim}{>}}\,}

\makeatletter

\let\@internalcite\cite
\def\cite{\@ifstar{\citeyear}{\citefull}}
\def\citefull{\def\astroncite##1##2{##1 ##2}\@internalcite}
\def\citeyear{\def\astroncite##1##2{##2}\@internalcite}

\def\citeau{\def\astroncite##1##2{##1}\@internalcite}
\def\citen{\def\astroncite##1##2{##1 (##2)}\@internalcite}
\def\possesivcite{\def\astroncite##1##2{##1's (##2)}\@internalcite}

\def\@citex[#1]#2{\if@filesw\immediate\write\@auxout{\string\citation{#2}}\fi
  \def\@citea{}\@cite{\@for\@citeb:=#2\do
    {\@citea\def\@citea{; }\@ifundefined
       {b@\@citeb}{{\bf ?}\@warning
       {Citation `\@citeb' on page \thepage \space undefined}}%
{\csname b@\@citeb\endcsname}}}{#1}}
\def\@cite#1#2{#1\if@tempswa , #2\fi}
\def\@biblabel#1{}
\makeatother

\begin{document}

\title{On  the  formation of  oxygen--neon  white dwarfs in close binary
       systems}

\author{Pilar Gil--Pons\inst{1} \and 
        Enrique Garc\'{\i}a--Berro\inst{1,2}}

\titlerunning{Formation of ONe white dwarfs}

\institute{$^1$Departament   de   F\'\i   sica   Aplicada,   Universitat
	       Polit\`ecnica  de Catalunya,  c/Jordi Girona s/n, M\'odul
	       B-4,  Campus  Nord,  08034  Barcelona,   Spain,  (e-mail:
	       pilar, garcia@fa.upc.es)\\
	   $^2$Institute for Space Studies of Catalonia, c/Gran Capit\'a
	       2--4, Edif.  Nexus 104, 08034 Barcelona, Spain}

\date{Received 22 February 2001 / Accepted 25 May 2001}

\abstract{
The evolution of a star of initial mass 10 $M_{\odot}$,  and metallicity
$Z = 0.02$ in a Close  Binary  System  (CBS) is  followed  from its main
sequence until an ONe degenerate  remnant forms.  Restrictions have been
made on the  characteristics  of the companion as well as on the initial
orbital  parameters  in order to avoid the  occurrence  of reversal mass
transfer  before  carbon is ignited in the core.  The  system  undergoes
three mass loss  episodes.  The first and second ones are a  consequence
of a case B Roche lobe  overflow.  During  the third  mass loss  episode
stellar winds may play a role comparable to, or even more important than
Roche lobe overflow.  In this paper, we extend the  previously  existing
calculations  of stars of  intermediate  mass  belonging to close binary
systems by following  carefully the carbon  burning phase of the primary
component.  We also propose different possible outcomes for our scenario
and discuss the  relevance  of our  findings.  In  particular,  our main
result  is  that  the   resulting   white   dwarf   component   of  mass
$1.1\,M_\odot$  more  likely  has a core  composed  of oxygen  and neon,
surrounded  by a mantle of  carbon--oxygen  rich  material.  The average
abundances  of the  oxygen--neon  rich core are $X({\rm  O}^{16})=0.55$,
$X({\rm    Ne}^{20})=0.28$,    $X({\rm   Na}^{23})=0.06$   and   $X({\rm
Mg}^{24})=0.05$.  This  result  has  important   consequences   for  the
Accretion  Induced  Collapse  scenario.  The average  abundances  of the
carbon--oxygen  rich  mantle are  $X({\rm  O}^{16})=0.55$,  and  $X({\rm
C}^{12})=0.43$.  The   existence  of  this  mantle  could  also  play  a
significant role in our understanding of cataclysmic variables.
}

\maketitle

%_______________________________________________________________________

\section{Introduction}

Intermediate  mass close  binaries are defined as those systems in which
the primary component develops a degenerate  carbon--oxygen  core, after
burning central helium in non--degenerate  conditions.  From the orbital
parameters  in these  systems, we see that periods are small enough that
the  possibility of mass transfer due to Roche Lobe overflow is enabled.
The evolution of low-- to  intermediate--mass  stars  belonging to close
binary  systems has been  widely  studied so far and, even  though  many
questions still remain unsolved,  important  contributions  have already
been  made  on  this  subject.  One  of  these  questions  concerns  the
evolution  of  heavy--weight  intermediate  mass stars (that is, primary
stars with masses between $\sim 8$ and 11 $M_\odot$)  belonging to close
binary  systems.  For the case of isolated  stars, this range of stellar
masses  corresponds  to stars for  which,  after  exhaustion  of central
helium,  carbon is  ignited  under  conditions  of  partial  degeneracy.
Ultimately, these stars become Super--AGB stars with ONe cores (Ritossa,
Garc\'\i a--Berro \& Iben 1996).  For the case of stars within this mass
range belonging to binary systems, very few comments can be made, either
because most of the calculations do not follow the evolution through the
carbon burning phase or, simply, because the existing calculations focus
mostly on a lower segment of masses.

For instance, Whyte \& Eggleton (1980) studied the evolution of stars of
up to 3 $M_{\odot}$  belonging to  semidetached  systems.  These authors
later extended their work to more general  scenarios in which  accretion
and mass transfer between low mass contact binaries were included (Whyte
\& Eggleton  1985).  Van der Linden (1987) also  performed  conservative
Case B  evolutionary  calculations  for several  masses of the  primary,
ranging from 3 $M_\odot$ to 12 $M_\odot$, but the evolution  through the
carbon  burning  phase  was not  followed.  Besides  the work  they have
performed  in the  field of  massive  binaries,  de Loore \&  Vanbeveren
(1992, 1994, and references  therein) have also focused on the evolution
of intermediate mass close binary systems (de Loore \& Vanbeveren 1995).
However, they were only able to follow the evolution of the primary star
until the  exhaustion  of the  helium in the  core.  Their  calculations
included  both  non--conservative  (de  Loore  \&  de  Greve  1992)  and
conservative   mass  transfer  (de  Loore  \&  Vanbeveren   1995).  Very
recently, Nelson \& Eggleton (2001) have performed a very  comprehensive
work  on  intermediate  mass  close  binary  systems,  exploring  5\,500
evolutionary tracks of mostly Case A conservative mass transfer systems.
The upper mass limit in this case was $\sim 50$ $M_\odot$ but, again, in
most of the cases the evolution  during the carbon burning phase was not
followed or it was aborted earlier (when the carbon luminosity  exceeded
1  $L_\odot$).  In  another  recent  work Han, Tout \&  Eggleton  (2000)
determined  the final mass relation for binary  systems with the mass of
the components  ranging between 3 and 8 $M_{\odot}$,  starting mass loss
at different times of the  Hertzsprung--Russell  gap.  However, in these
studies the  assumption of  conservative  mass  transfer was adopted and
justified by the conditions of the case they consider which, in spite of
corresponding to an important  portion of the real cases, cannot account
for all of them.

Iben (1985,  1991) has  extensively  reviewed  the  physical  mechanisms
relevant to binary systems and has thoroughly discussed the evolution of
intermediate mass close binary systems,  offering an excellent  overview
of their evolution until very late stages, proposing  several  different
scenarios and providing their  probabilities  of occurrence.  Also, Iben
\& Tutukov  (1984) have proposed  different  evolutionary  scenarios for
heavy--weight  intermediate  mass close binary systems as progenitors of
SNe Ia.  In spite of the fact that this mass  interval  contains  a good
fraction of the stars  which are  massive  enough to ignite  carbon in a
non--explosive  way, the  evolution  of  these  systems  has  been  much
neglected  until very recently.  The  pioneering  works of Miyaji et al.
(1980) and Woosley,  Weaver \& Taam (1980) lead to the  conclusion  that
the stars of this mass interval  belonging to close binary systems would
lose   most   of   their    mass   and,    moreover,    would    develop
electron--degenerate  ONe cores  after the carbon  burning  phase.  In a
second phase, due to accretion from the secondary,  the central  density
would increase until the threshold for electron capture on $^{24}$Mg and
$^{24}$Na  would be reached first, and on $^{20}$Ne and $^{20}$F  later.
The process of electron  capture on these nuclei would  finally  trigger
the  explosive  ignition  of neon  and  oxygen  at very  high  densities
($\rho_{\rm c} \ga 2 \times  10^{10}$~g~cm$^{-3}$).  At these densities,
fast  electron  captures on the  incinerated  material  would  bring the
Chandrasekhar  mass  below the  actual  mass of the ONe core and  induce
gravitational  collapse to neutron star dimensions.  Although there is a
general  agreement  that  electron--capture  induced  collapse  could be
successful,  a major  drawback  of this  scenario  is that  no  detailed
pre--collapse  models existed in the  literature.  For instance, in most
of the  calculations,  the evolution  during the mass loss phase was not
followed in full detail (Nomoto 1984) or the evolution during the carbon
burning phase was  oversimplified  by introducing the so--called  steady
burning  approximation  (Saio \& Nomoto 1998).  Therefore,  the depicted
evolutionary  scenario  could  be  substantially  modified  due  to  the
presence of these approximations.

Very  recently, the evolution  leading to the  formation of white dwarfs
with ONe cores in close  binary  systems  is being  reinvestigated.  For
instance,  Dom\'\i  nguez,  Tornamb\'e  \& Isern (1994) have studied the
formation of an ONe white dwarf through mass  transfer in a close binary
system.  On the  other  hand, in a series  of  recent  papers  (Garc\'\i
a--Berro \& Iben 1994; Ritossa, Garc\'\i a--Berro \& Iben 1996; Garc\'\i
a--Berro, Ritossa \& Iben 1997; Iben, Ritossa \& Garc\'\i a--Berro 1997;
Ritossa,  Garc\'\i  a--Berro  \& Iben 1999) the  evolution  of  isolated
heavy--weight  intermediate mass stars has also been carefully  followed
from the main  sequence  phase up to  exhaustion  of carbon in the core.
Perhaps one of the most  important  conclusions  of these papers is that
isolated white dwarfs with masses  $M_{\rm  WD}\ga  1.0\,M_\odot$  would
most likely have an ONe core  surrounded  by a mantle of  carbon--oxygen
rich   material.   This   bears   important    consequences    for   the
above--mentioned  accretion--induced  collapse  scenario because all the
existing  calculations  assume that the composition of the He--exhausted
core is  carbon--free.  Nevertheless,  these  authors  studied  only the
mass--conservative evolutionary tracks for the relevant range of stellar
masses,  whereas in a close binary system the  composition  of the final
remnant  could be  dramatically  affected by the  previous  evolutionary
phase  if the  star  is  interacting  with a  companion.  In  any  case,
progress in the right  direction has been made, but further  exploration
is still worthwhile.

\begin{figure*}
\vspace{14.5cm}
\hspace{1.0cm}
\includegraphics{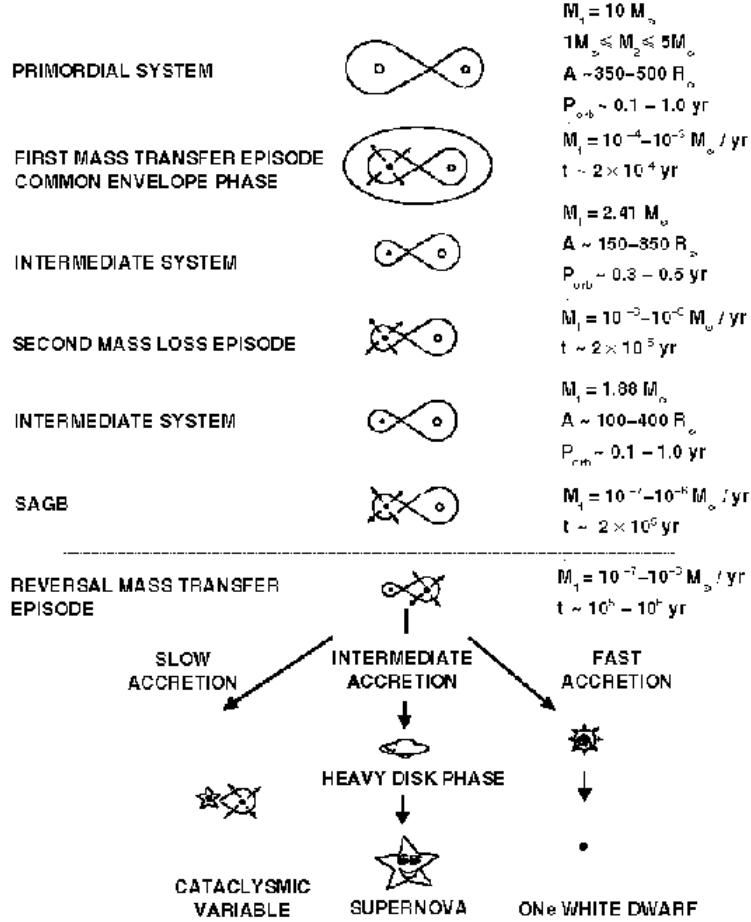}
\caption{Approximate  evolution of the orbital  parameters of the binary
system and  outline of the  possible  final  outcomes.  The dashed  line
divides the  calculations  reported in this paper and the three possible
outcomes.}
\end{figure*}

In this paper we follow the evolution of a 10 $M_{\odot}$  model star of
solar  metallicity  belonging to a close  binary  system,  from its main
sequence phase until an  oxygen--neon  core  develops.  In order to keep
consistency with our previous results, for the calculations  reported in
this paper we have used the same evolutionary code described in Ritossa,
Garc\'\i  a--Berro \& Iben (1996).  In particular we have not considered
overshooting,  as was done there.  This procedure might lead to somewhat
smaller cores than in the case in which  overshooting is considered, but
we expect the differences to be small.  Different mass loss episodes are
caused by the presence of a close  companion and are followed in detail.
Specifically, the model loses most of its  hydrogen--rich  envelope in a
case B Roche--lobe  overflow  episode.  It is worth  mentioning  at this
point that we have taken  special  care in treating  the mass zoning and
the time steps  during the mass loss  episodes  and the  carbon  burning
phase  (about  2\,000  lagrangian  mesh points are used  during the most
problematic  phases).  Moreover,  the  distribution  of mesh  points  is
regularly  updated at each time step.  Our  algorithm  puts mesh  points
where they are most needed (that is, where the gradients of the physical
variables are strong) and eliminates  them where they are not necessary.
Once this is done, the values of the physical variables are interpolated
at the new mesh  points  in  order  to  properly  compute  the  temporal
derivatives.  If necessary the mesh can be updated at each iteration.

As is usually done,  throughout this work we will refer to the star that
is initially more massive as the primary component whereas the secondary
component  will be its  companion.  We have  assumed  that  the  initial
orbital  parameters  and the mass of the secondary  component  allow the
whole  evolution to proceed  without  reversal mass transfer and without
disruption or merger events.  This poses some constraints on the mass of
the secondary which will be discussed later.

The plan of the paper is the  following.  In the second  section of this
paper we explain our choice of the initial  orbital  parameters  and the
assumed  scenario.  In the third section we present a description of the
overall  evolution of our model star until carbon  ignition sets in, and
we compare it with the  evolution of an isolated  star of the same mass.
This  section is also  devoted to the study of the  resulting  mass loss
episodes.  In the fourth  section  we  briefly  describe  the  evolution
during the carbon burning phase and we discuss the final characteristics
of  the  remnant.  Finally  our  major  findings  and  conclusions   are
described in section \S 5.

%_______________________________________________________________________

\section{The scenario}

In this section we set up our evolutionary  scenario and we describe our
choices for the initial  orbital  parameters.  The reader should keep in
mind that our main goal is to provide a successful  scenario to test the
formation of massive ONe white dwarfs in close binary systems that could
ultimately  lead either to a  cataclysmic  variable  or to  collapse  to
neutron  star  dimensions   through  the   accretion--induced   collapse
alternative.  Therefore  our  choice of the  initial  parameters  of the
binary system is effectively influenced by the desired final outcome.  A
possible observational counterpart of the proposed scenario could be the
binary system IK Peg (HR~8210,  HD~204188),  which has an orbital period
of 21.7  days,  and it is  composed  of a  massive  white  dwarf of mass
$M_{\rm WD} \simeq 1.15\,  M_\odot$ and a main  sequence  star of $1.4\,
M_{\odot}$ (Smalley et al., 1996).

Our starting point is a primordial  system  composed of a 10 $M_{\odot}$
star and its $\la 5\,  M_{\odot}$  companion,  which we will refer to as
the primary and secondary  components  respectively  (see  Fig.~1).  The
system  undergoes  a case B mass  transfer,  this being the most  likely
possibility,  and so the primary  starts  losing mass after the onset of
hydrogen  burning in a shell.  Unlike early case--B mass loss  episodes,
late  case--B  mass  transfer  has been  little  studied up to now.  The
reasons for this are multiple but perhaps the most  important one is the
additional  computational  difficulties  that arise when studying such a
phase.  However, it has been recently pointed out (Tauris \& Dewi, 2001)
that the actual  definition of the  resulting  core of the primary after
the mass loss episode could influence the final orbital  parameters.  We
consider it  interesting,  therefore,  to study a late case--B mass loss
episode.  This determines the range of values for the initial Roche lobe
radius of the  primary,  that we  actually  choose  by  considering  the
evolutionary  track for the single $10 \, M_{\odot}$ model star followed
by  Garc\'\i  a--Berro  et al.  (1994),  and  we  keep  it  constant  at
$210\,R_\odot$  during the whole  process  (see the  discussion  below).
Nonetheless, it is worth  mentioning  that we have conducted a series of
numerical experiments in which the Roche lobe radius has been changed to
values  as small as  $120\,R_\odot$  and we have not  found  significant
differences.  Given the initial mass  relation and the Roche lobe of the
primary, we can obtain the orbital separation between both components by
appliying the equation (Eggleton, 1983):

\begin{equation}  
\frac{R_{\rm L}}{A}= {\frac{0.46q^{2/3}}{0.6q^{2/3}+\ln{(1+q^{1/3}})}}, 
\end{equation}

\noindent  where $R_{\rm L}$ denotes the  effective  radius of the Roche
lobe of the primary, $A$ indicates the orbital separation and $q$ is the
mass ratio  ($M_1/M_2$)  between the  components.  The  initial  orbital
separation  turns out to be  $A\sim  150-350\,R_\odot$  and the  orbital
period is $P_{\rm orb}\sim 0.1-1.0$~yr, depending on the initial mass of
the  companion  (between  1 and  5~$M_{\odot}$),  which  are  reasonable
values.

As it will  be  shown  in  detail  in the  next  section,  a  number  of
uncertainties  are  involved  in the  first  mass  loss  episode.  These
uncertainties  are  basically  caused  by  the  formation  of  a  common
envelope.  This common  envelope is formed when the primary star attains
the highest values for its mass loss rates, and most probably  removes a
large   amount  of  mass  and   angular   momentum   from  the   system.
Consequently,  the  evolution  of  the  orbital   parameters  cannot  be
accurately  followed, and this can only be done in an  approximate  way.
In order to determine  the mass lost by the  primary, we assume that all
the matter overflowing the Roche lobe is lost by the star, and we simply
stop the process when its radius  dramatically  decreases  below that of
the Roche lobe (see \S 3.1).

The second  mass loss  episode  starts  when the  surface  radius of the
primary again exceeds that of its Roche lobe.  In this case, this is due
to the onset of  helium  burning  in the  shell.  This  process  is most
likely  conservative  both in mass and in angular  momentum  and, again,
stops when the radius of the  primary  decreases  below the value of the
Roche lobe.  After this temporary  decrease in radius,  carbon  ignition
continues  and the primary  starts the ascent along the  Super--AGB  but
does not fill again its Roche  lobe.  Once  carbon is  exhausted  in the
innermost  regions, it is a stellar wind rather than Roche lobe overflow
that  induces  mass  loss  and,  thus,  we use  the  parametrization  of
Nieuwenheuzen  and de Jager (1990) in order to account for the mass loss
rates.  As  happened  during  the first  mass  loss  episode,  there are
several   uncertainties   involved  in  this  process,  which  can  have
consequences on the final fate of the system.  In fact, depending on the
choice of the efficiency of the winds, the resulting orbital  parameters
can vary significantly (Umeda et al.  1999).  If the system survives the
Super--AGB  phase, we are left with an ONe white  dwarf,  as a result of
the  evolution of the primary,  and a main  sequence  star that, at some
time, will fill its Roche lobe and give rise to reversal mass  transfer.

Even though we do not mean to get deep into the study of the probability
of occurrence of the different final  outcomes, we will briefly  outline
the  different  possibilities  for the final  stages  of the life of the
binary in terms of the mass  transfer  rates  during the  reversal  mass
transfer.  Depending  on this mass  transfer,  the final  outcome can be
either a  cataclysmic  variable  if it is lower than a critical  rate, a
supernova  explosion if it is larger or an ONe white dwarf if it is much
larger.  A more detailed study can be found in \S 5.

%_______________________________________________________________________

\section{Evolution  until the beginning of the carbon  burning phase and
main mass loss episodes}

\begin{table*} 
\hspace{3cm}
\begin{center}
\begin{tabular}{cccccccccccc}
\multicolumn{9}{c}{Table  1.  ~~  Characteristics  of models at  various
                   points in the H--R diagram of Fig.  2.}  \\
\hline
Model & $t\, (10^{14}{\rm s})$ & $\log L$ 
      & $\log T_{\rm eff}$     & $\log R$ 
      & $\log\rho_{\rm c}$     & $\log T_{\rm c}$ 
      & $M_{\rm He}$           & $M_{\rm C}$
      & $M_{\rm 1}$            & $M_{\rm 2}$ 
      & $P_{\rm orb}$ (yr) \\
\hline
A      &    6.1785 & 3.79 & 3.67 & 2.07 & 6.67 & 8.16 & 1.97 & 0.00 & 10.0 & $1.0-5.0$ & $0.1-1.0$ \\
B      &    6.1795 & 4.12 & 3.63 & 2.32 & 6.62 & 8.16 & 1.97 & 0.00 &      &           &           \\
C      &    6.1893 & 4.16 & 3.64 & 2.32 & 6.51 & 8.16 & 1.97 & 0.00 &  2.4 & $1.7-7.0$ & $0.3-0.5$ \\
D      &    6.2917 & 3.99 & 3.64 & 2.23 & 6.45 & 8.18 & 2.04 & 0.00 &      &           &           \\
E      &    6.2936 & 3.96 & 3.59 & 2.32 & 6.45 & 8.18 & 2.04 & 0.00 &      &           &           \\
F      &    6.2954 & 3.98 & 3.59 & 2.32 & 6.44 & 8.18 & 2.04 & 0.00 &      &           &           \\
G      &    6.3555 & 3.84 & 3.62 & 2.41 & 6.43 & 8.19 & 2.05 & 0.00 &      &           &           \\
H      &    6.8505 & 3.68 & 4.18 & 2.01 & 6.47 & 8.24 & 2.05 & 0.76 &      &           &           \\
I      &    7.2076 & 3.90 & 3.72 & 2.02 & 7.42 & 8.15 & 2.05 & 0.86 &      &           &           \\
J      &    7.2122 & 3.82 & 3.81 & 1.85 & 7.43 & 8.42 & 2.05 & 0.91 &      &           &           \\
K      &    7.2178 & 3.89 & 3.64 & 2.18 & 6.43 & 8.45 & 2.06 & 0.93 &      &           &           \\
L      &    7.2638 & 4.21 & 3.73 & 2.18 & 7.66 & 8.60 & 2.06 & 1.04 &  1.9 & $2.2-9.3$ & $0.1-1.0$ \\ 
\hline
\end{tabular}
\end{center}
\end{table*}

%
% TABLA DE ABUNDANCIAS SUPERFICIALES TRAS EL PRIMER DREDGE UP
%
\begin{table*}
\hspace{3cm}
\begin{center}
\begin{tabular}{lcccccc}
\multicolumn{7}{c}{Table  2.  ~~  Surface  abundances  after  the  first
		   dredge--up.  SS stands for single star.}  \\
\hline
Model & $X_{\rm H}$      & $X_{\rm 3}$
      & $X_{\rm He}$     & $X_{\rm C}$
      & $X_{\rm N}$      & $X_{\rm O}$ \\
\hline
10 $M_{\odot}$ (SS)   &  0.681 & $1.23 \times 10^{-5}$ & 0.305 &
                         $1.79 \times 10^{-3}$ & $2.83 \times 10^{-3}$
                         & $7.71 \times 10^{-3}$ \\
10 $M_{\odot}$ (CBS)  &  0.664 & $1.13 \times 10^{-5}$ & 0.324 &
                         $1.64 \times 10^{-3}$ & $3.46 \times 10^{-4}$
                         & $7.19 \times 10^{-3}$ \\
9 $M_{\odot}$  (SS)   &  0.696 & $1.49 \times 10^{-5}$ & 0.291 &
                         $1.85 \times 10^{-3}$ & $2.48 \times 10^{-3}$
                         & $8.03 \times 10^{-3}$ \\
\hline
\end{tabular}
\end{center}
\end{table*}

The  presence of a close  companion  has  several  consequences  for the
evolution of our 10 $M_{\odot}$  model star.  In  particular,  there are
two Roche--lobe  overflow mass loss episodes.  The first one occurs just
after the main  sequence  phase when our model  star  reaches  red giant
dimensions,  and the second one happens  shortly after the exhaustion of
central helium.

Figure 2 shows the  evolution  of our model in the  Hertzsprung--Russell
diagram.  Times to evolve to each labeled  point along the  evolutionary
track are given in the second  column of Table 1, where we also  provide
the most important  characteristics  of these particular  models for the
hydrogen and helium  burning  phases.  Also shown in the last columns of
table 1 are the mass of the  primary  and  secondary  stars  ($M_1$  and
$M_2$), and the  expected  orbital  period  $(P_{\rm  orb})$.  The solid
lines  represent  the  evolutionary  phases  during  which no mass  loss
occurs, whereas the dotted lines correspond to the  evolutionary  phases
where mass loss  occurs:  from B to C, for the first mass loss  episode,
and  from  shortly  after K to L, for  most  of the  second  Roche--lobe
overflow.  Note, however, that the final part of the second  Roche--lobe
overflow occurs when carbon has already been ignited in the core (see \S
3.4 and \S 4) and, thus, it is not shown in figure 2.  As expected,  the
effects of the mass loss episodes  considerably  modify the evolutionary
track when compared with the  evolution of an isolated  star of the same
initial mass (Garc\'\i  a--Berro \& Iben 1994).  For instance,  although
the descent along the red giant branch takes place at a slightly  higher
temperature  than in the case of an isolated star  (4\,400~K  instead of
4\,300~K),  it is  followed  by a sudden  shift to bluer  regions of the
diagram due to the adiabatic  expansion and cooling that accompanies the
first mass loss episode (see below).  The second major different feature
is that the developement of the blue loop during the core helium burning
phase takes place at lower luminosities than in the case of the isolated
model   star   ($6.9\times    10^3\,L_\odot$   instead   of   $7.9\times
10^3\,L_\odot$).

\begin{figure}
\vspace{7.9cm}
\hspace{-2.7cm}
\includegraphics{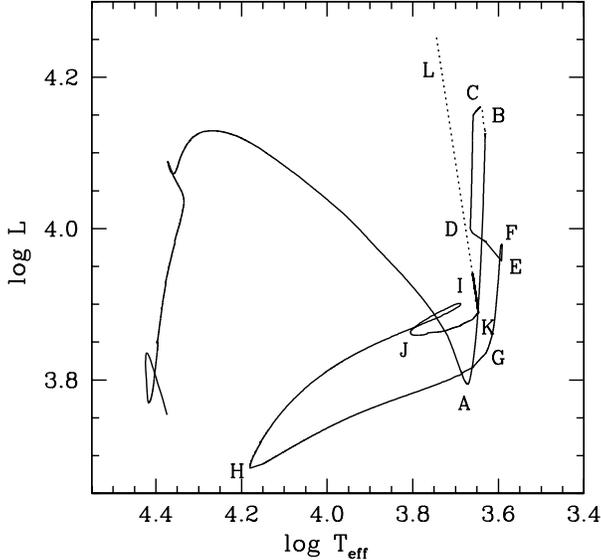}
\caption[]{Evolutionary   track  of  the   primary   component   in  the
Hertzsprung--Russell  diagram.  The  physical  quantities  of the labeled
models are shown in Table 1.}
\end{figure} 
	
\begin{figure}
\vspace{7.9cm}
\hspace{-2.7cm}
\includegraphics{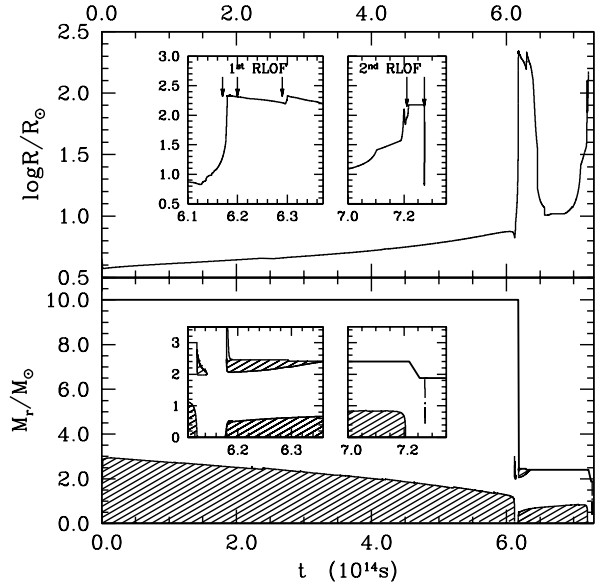}
\caption[]{Evolution  of the main structure  parameters as a function of
time  during the first and the  second  mass--loss  episodes.  The upper
panel shows the  evolution  of the radius of the  primary.  The  inserts
show the  mass--loss  episodes with higher  resolution.  The lower panel
shows  the  convective  zones  engendered  during  the  evolution  up to
off--center ignition of carbon.}
\end{figure}

In Figure 3 we show the  temporal  evolution  of the  radius of the star
(top  panel)  and of the  convective  regions  associated  with  nuclear
burning  (bottom  panel),  from the zero  age  main  sequence  up to the
off--center  ignition  of  carbon  in the  helium--exhausted  core.  The
inserts show with higher  resolution  the  evolution  of the  convective
regions  and of the  radius  during  the two  mass--loss  episodes.  The
initial  central  convective  zone of Fig.  3 is due to the high  fluxes
engendered  by the  CN--cycle  reactions  and  persists  until  hydrogen
vanishes at the center.  An off--center  convective  region forms later,
very  much  in the  same  way as in the  case  of a  single  star.  This
convective  region  is due to the  release  of  gravitational  potential
energy during the overall  contraction phase that follows the exhaustion
of central  hydrogen  and the  establishement  of the  hydrogen--burning
shell that  occurs from points C to D in Fig.  2.  The thick  solid line
in the lower panel  corresponds  to the total mass of the primary which,
as it can be seen, decreases  dramatically  at the begining of the first
dredge--up--process,   which  occurs   simultaneously   with  the  first
Roche--lobe  overflow.  Since  this  mass  loss  episode  occurs  in the
presence of a deep convective  envelope, the associated  time scales are
short.  On the contrary, we will see that the second mass--loss  episode
is  much  more  stable  since  it is not  associated  with a  dredge--up
episode.  This behaviour  makes the two mass--loss  episodes  completely
different and has  important  consequences.  For  instance, we expect to
find  a  different  pattern  of  surface  composition  after  the  first
mass--loss  episode and dredge--up,  when compared to the evolution of a
single  star.  The surface  composition  of both model stars and that of
the  isolated  9  $M_\odot$  star just  after the end of the  dredge--up
episode can be found in Table 2.  As we shall show below, the comparison
with the 9 $M_\odot$  star is relevant  for this study since some of the
results  obtained in the  calculations  reported here are much closer to
the isolated 9 $M_\odot$ model than to those of the 10 $M_\odot$  single
star.  As  can  be  seen,  the  helium   and   nitrogen   contents   are
significantly  higher in the case of a model star  belonging  to a close
binary system.

%
% THE FIRST MASS LOSS EPISODE
%

\subsection{The first mass loss episode}

Our model undergoes a Case B mass loss process that starts shortly after
hydrogen  combustion  in a shell has been  established,  and the surface
radius reaches the Roche Lobe radius which, as previously  mentioned, we
have  adopted  to be  210  $R_{\odot}$.  In  computing  this  mass--loss
episode we have (somewhat  arbitrarily) assumed that the primary keeps a
constant radius which is coincidential  with the Roche--lobe  radius and
that all the  overflowing  matter will be lost by the primary.  Although
this  is  a  classical   prescription   there  are,  of  course,   other
alternatives  (Nelson  \&  Eggleton  2000).  It  is  nevertheless  worth
noticing that when using the last approximation with secondary masses of
around  $\sim\,5\,M_\odot$, the Roche lobe radius changes by $\sim 10\%$
and,  consequently,  we do not expect this to have a large impact on our
results.  We thus defer such study to a forthcoming publication.

\begin{figure}[ht]
\vspace{7.8cm}
\hspace{-2.9cm}
\includegraphics{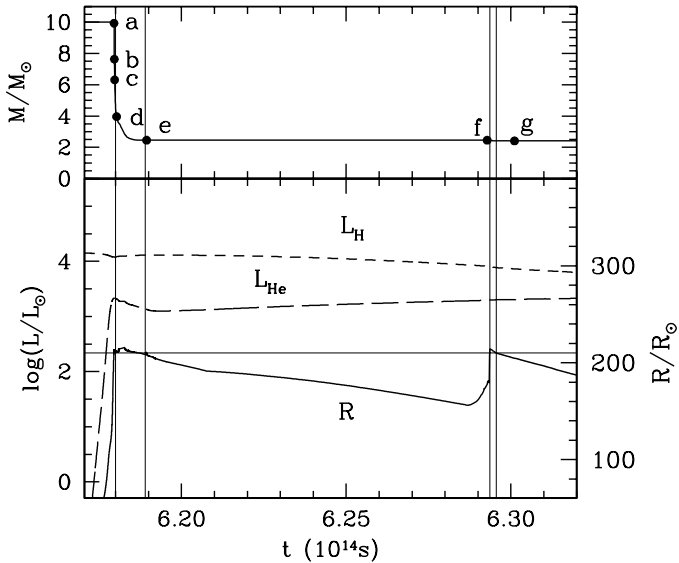}
\caption[]{Global characteristics of the primary component as a function
of time during the first mass--loss  episode.  The upper panel shows the
total mass as a function  of time.  The lower  panel  shows the  surface
radius and the  luminosities  provided by hydrogen  and helium  burning.
The  vertical  thin  lines  delimit  the two  phases of mass  loss.  The
surface luminosity is mostly provided by hydrogen combustion.}
\end{figure}

In the first  insert in the top and  bottom  panels of Fig.  3 the first
mass--loss episode is shown.  Also, in Figure 4 we show the evolution of
several  interesting  quantities  like the hydrogen  and helium  nuclear
luminosities.  This  mass--loss  episode occurs in two distinct  phases.
The first and most violent  phase occurs as the  convective  envelope is
still  advancing to the interior of the star and when helium has already
been  ignited  at the  center,  leading  to the  formation  of a central
convective region, which reaches its equilibrium value when $\log(L_{\rm
He}/L_\odot)\simeq 3.3$.  This phase is clearly marked by the two arrows
in the top insert of Fig.  3 and by the thin vertical  lines of Fig.  4.
During this first phase, the bulk of the hydrogen--rich envelope is lost
($\sim 7.6 \,  M_\odot$),  leaving a remnant of 2.4  $M_\odot$  of which
$\sim 0.5\, M_\odot$  corresponds  to the remaining  H--rich  convective
envelope.  The duration of this phase is $3.0 \times 10^4$ yr.

Since the dependence of the variation of the total radius of the star on
the mass is  $\Delta R \sim  RM^{-1/3}$  (de  Loore \& Doom  1992),  the
mass--loss process  experiences a positive  feedback.  Therefore, during
the very first  part of this  mass--loss  episode  the  feedback  of the
process  allows very high values for the mass loss rates, that can reach
values as high as $\dot{M_1}  \sim 10^{-3} \, M_{\odot} {\rm  yr}^{-1}$.
At these very high mass--loss rates the star is no longer able to keep a
constant  radius and at the end of this phase the radius of the  primary
falls  below the value of the  Roche--lobe,  and thus the mass  transfer
onto the secondary  temporarily stops.  As the evolution  continues, the
primary  again fills its  Roche--lobe  radius,  leading to a  subsequent
phase of mass--loss.  During this second phase only $\sim 0.05\,M_\odot$
are lost by the primary.  Thus, a small portion of the H--rich  envelope
remains even at the end of the first mass--loss episode.  The time scale
for this second phase is  significantly  longer ($1.6 \times  10^5$ yr),
leading to much more modest  mass--loss  rates:  $\dot{M_1} \sim 10^{-5}
\, M_{\odot} {\rm yr}^{-1}$.

The existence of a deep convective envelope surrounding the H--exhausted
core of the primary and,  consequently, the high values of the resulting
mass--loss  rates very much enhance the  possibility  that the system is
embedded in a common  envelope.  No definite and accurate  study of this
kind of structure  has been  performed up to now.  Thus, the part of the
process  in  which  a  common   envelope  forms  is  plagued  with  many
uncertainties, the most important one perhaps being the influence of the
common envelope on the orbital parameters.  The expressions  provided by
Eggleton   (1983)  and   Paczy\'nski   (1971)   can  only  be  taken  as
approximations,  or even upper  limits, and it is  necessary  to rely on
estimates that have been obtained for other systems that presumably have
undergone a common envelope phase, like the cataclysmic  variable U~Gem.
These estimates yield values for the angular momentum losses that can be
as high as 70--90\% (Iben \& Tutukov  1984).  In any case, when a common
envelope is present, the angular momentum losses are expected to be high
due to the release as sound waves of the energy  generated by frictional
drag  in  the  matter  of  the  envelope.  Therefore,  from  an  initial
separation   between  the  components  of  $A_{\rm  i}\sim   350-450  \,
R_{\odot}$  (correspondig  to $1  \,M_{\odot}  \leq  M_{\rm  2}  \leq  5
\,M_{\odot}$  respectively)  one  can  assume  a  typical  80\%  angular
momentum loss and apply the equation

\begin{equation}
J_{\rm orb}^2 = GA_{\rm i} \frac{(M_1 M_2)^2} {M_1 + M_2}
\end{equation}

\noindent to get the final separation between components after the first
Roche--lobe overflow, $A_{\rm f} \sim\, 150-350\, R_{\odot}$.  It should
be noted,  nevertheless,  that this way of estimating  the final orbital
parameters  usually leads to larger orbital  separations than the values
obtained when using the treatment of Webbink (1984).

We can estimate the ratio between the mass accreted by the secondary and
the mass lost by the  primary, $ \beta$, by  assuming  stable  accretion
onto the  secondary.  Thus,  ${\dot M_2} = M_2 / \tau_{\rm  KH}$,  where
$\tau_{\rm KH}$ is the Kelvin--Helmholtz timescale.  Considering typical
values for the luminosity and for the surface  radius of a main sequence
companion of mass between 1 and 5 $M_{\odot}$, we get a reasonable value
for the mass  accretion  rate of the  secondary, $ \dot M_2 \sim 10^{-4}
M_{\rm _{\odot}}$.  Taking into account that most of the mass is lost by
the  primary  during  the first  $3.0  \times  10^4$  yr, we can  easily
calculate  an {\sl  average}  value for  $\beta$  during  this mass loss
episode,  which turns out to be  $\beta\sim  0.1$.  Of course, it can be
argued as well that  another  reasonable  way to estimate  $\beta$ is to
adopt the Eddington limit instead of the thermal timescale, this being a
firm upper limit.  For the range of relevant  parameters, this procedure
would  lead  in  our  case  to  accretion   rates  3  times  larger,  or
equivalently, to $\beta\sim  0.3$.  Nevertheless it should be noted that
the  calculations  of Hjellming \& Taam (1991)  predict a  significantly
lower value of $\beta$,  of the order of 0.01.  Thus, the mass  accreted
by the  secondary  will be in the  range  $0.7-2.0\,M_\odot$,  at  most.
Since the main  sequence  lifetime of a $7.0 \, M_\odot$  star is $t\sim
9.0\times 10^{14}$~s it is clear that reversal mass transfer will not be
enabled  until carbon is exhausted in the inner core of the primary (see
Table 1 and \S 4 below), in accordance with our scenario.

\begin{figure}[ht]
\vspace{7.8cm}
\hspace{-2.9cm}
\includegraphics{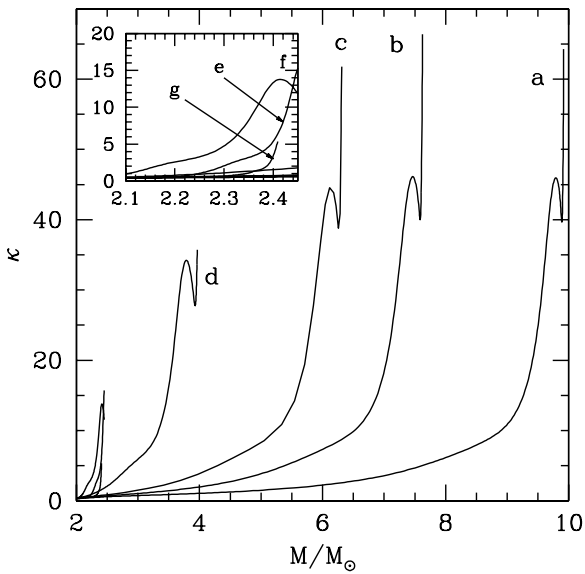}
\caption[]{Opacity  profiles in the  hydrogen--rich  envelope during the
first  Roche lobe  overflow.  The  models  displayed  correspond  to the
different  times  labeled in Fig.  4.  The  insert  shows,  with  higher
resolution, the opacity profiles for the last models.}
\end{figure}

We have  considered the  possibility  that part of the matter ejected by
the primary component might be accreted back by the same star.  In order
to obtain some hints of the possible  outcomes to this  problem, we have
estimated  the  duration  of the  common  envelope  phase  as in Iben \&
Tutukov  (1984)  and  compared  it with the  thermal  time  scale of the
primary.  As the latter is much longer  than the  expected  time for the
common  envelope to remain bound to the system, there are strong reasons
to admit that there is not enough time for the ejected gas to cool down,
lose kinetic energy and be overtaken by the  gravitational  potential of
the primary.

It is also  worth  noting  that the  duration  of the  first  dredge--up
episode is longer in the calculations  reported here than in the case of
the isolated model described in Garc\'\i  a--Berro \& Iben (1994).  This
is due to the fact that the fast release of gravitational  energy during
the compression  phase that happens in the middle of the first mass loss
episode  cannot be  evacuated  solely by  radiation  and,  thus,  allows
convection  to  persist.  On the other  hand, one should not forget that
the overall  evolutionary  time scales are longer for decreasing  masses
and this effect  would also have an  influence  on the  duration  of the
dredge--up episode.

%
% BEHAVIOUR OF THE ENVELOPE DURING FIRST MASS LOSS EPISODE
%

In order to find an explanation  that, at least partially,  accounts for
the  behaviour of the  envelope  during the first mass loss  episode, in
Fig.  5 we show the opacity profiles of several  specific models for the
times  labeled  in Fig.  4  (models  {\tt a} to {\tt  g}).  The  fastest
expansion  phase is  coincidential  with  models  {\tt a} to {\tt c}, in
which the existence of a deep convective envelope, basically composed of
hydrogen at  relatively  low  temperature,  leads to a high opacity and,
thus,  to  an  inefficient   transport  of  the  energy.  Therefore,  an
important  portion of the energy generated at the hydrogen burning shell
is not driven  outwards but, instead, it is  transformed  into  internal
energy in the envelope and, then, into work of  expansion,  thus keeping
the value of the surface radius very close to the Roche lobe radius.

When the model reaches  approximately  4 $M_{\odot}$  (model {\tt d}), a
large portion of the  hydrogen--rich  envelope is already  lost, and the
inner and hotter  layers are exposed.  At this point, on the one hand we
have less mass able to absorb the flow of energy  and, on the other, the
opacity is also smaller.  Consequently,  both  phenomena  allow  nuclear
energy to flow almost freely to the surface  without  being  transformed
into work of expansion  and, at model {\tt e}, when the mass of the star
is about 2.45 $M_{\odot}$, a fast overall (almost adiabatic) contraction
of the convective  envelope  occurs,  leading the surface radius to drop
below the value of the Roche  lobe and,  hence,  mass  loss  temporarily
halts.  Finally,  for  model  {\tt  f}  the  opacity  increases  due  to
compression and, thus, a new phase of expansion  occurs which drives the
surface  radius to again reach the value of the Roche lobe radius.  Thus
the mass loss episode is restored for a brief interval,  until some more
cool  hydrogen--rich  layers are lost.  Finally the  opacity  definitely
decreases (model {\tt g}) and the mass loss episode is finished.

It is worth  mentioning  at this point that we get a remnant of slightly
higher  mass than that of the models  found in the  existing  literature
(Iben \& Tutukov 1985; Dom\'\i  nguez,  Tornamb\'e \& Isern 1994).  This
is due to the fact that in our model the hydrogen--rich  envelope is not
completely lost during the first Roche lobe overflow  episode.  However,
since the rest of the  remaining  hydrogen--rich  envelope is lost later
during the second Roche lobe  overflow  episode, and since the growth of
the helium core is very small during the time  between the two mass loss
episodes  (approximately 0.08 $M_{\odot}$), we expect that the influence
on the CO and ONe core sizes and  compositions  is negligible and so the
final results will not be substantially different.

%
% SECOND MASS LOSS EPISODE
%

\subsection{The second mass loss episode}

\begin{figure}[ht]
\vspace{7.8cm}
\hspace{-2.9cm}
\includegraphics{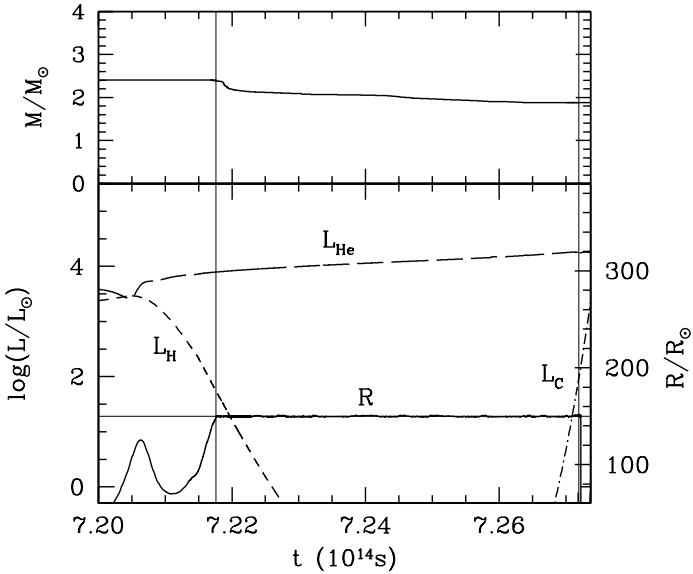}
\caption[]{Global characteristics of the primary component as a function
of time during the second mass loss  episode.  The upper panel shows the
total mass as a function  of time.  The lower  panel  shows the  surface
radius and the  luminosities  provided  by  hydrogen,  helium and carbon
burning.  The vertical thin lines delimit the second mass loss  episode.
The surface luminosity is mostly provided by helium combustion.}
\end{figure}

The second mass loss  episode  starts when  helium  burning  begins in a
shell and, as a consequence, its surface radius approaches again that of
the Roche  lobe (see Fig.  3), which  now is $\sim  80\,  R_\odot$.  The
global  characteristics  (the  mass  and  radius  and  the  luminosities
associated  with the active burning  regions) of the primary  during the
second  mass loss  episode  are shown in Fig.  6.  The fact  that in the
outer envelope, from which the matter is removed,  energy is transported
by radiation  instead of being  transported  by  convection,  allows the
process to take place in a more  stable way than in the first  mass loss
episode.  Hence, the mass loss rates are much  smaller  ($\dot  M_1 \sim
10^{-6} \, M_\odot {\rm yr}^{-1}$) during most of the process, except at
two phases,  during which the values of the mass loss rate are one order
of magnitude  higher.  The first phase  corresponds  to the beginning of
the mass loss episode, when the remainder of the hydrogen--rich envelope
is lost, and the second phase happens  nearly at the end of the process,
when  carbon  is  ignited   off--center.  The  end  of  the  process  is
determined  by a steep  decrease in the surface  radius of the star that
takes values below the Roche lobe radius.

The moderate values we get for the mass loss of the primary  support the
hypothesis of conservative  mass transfer.  Furthermore, one can compare
the luminosity  associated  with accretion onto the secondary,  which is
given  by  $L_{\rm  acc}  =  \dot   M_2(\Phi_{L_1}-\Phi_{R_2})$,   where
$\Phi_{L_1}$  and  $\Phi_{R_2}$  stand for the  gravitational  potential
computed  at the Roche lobe  radius and at the  radius of the  secondary
respectively,  with the Eddington  luminosity of the secondary,  $L_{\rm
Edd}$.  For a set of typical values for the secondary  star, we get that
$L_{\rm acc} \ll L_{\rm Edd}$ and, thus, according to Han et al.  (1999)
conservative mass transfer is most likely.

%
%   EVOLUTION OF CENTRAL STRUCTURE PARAMETERS.
%

\subsection{Overall characteristics before carbon ignition}

\begin{figure}[ht]
\vspace{7.8cm}
\hspace{-2.5cm}
\includegraphics{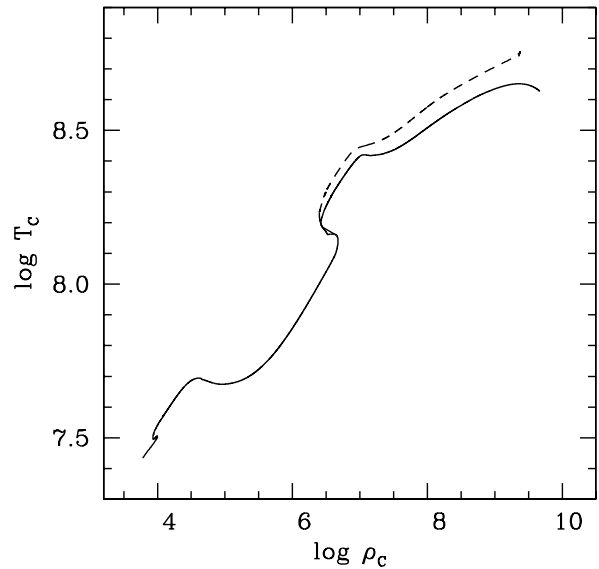}
\caption[]{Evolution  in the  $\log\rho  - \log T$ plane of the  central
layer of the isolated 10  $M_{\odot}$  star (dashed line) and the of the
primary  component of the close binary system studied here (solid line),
until the onset of carbon burning.}
\end{figure}

In Fig.  7 we show the  evolution  of the  center of the  primary in the
$\log\rho - \log T$ diagram  from the main  sequence  until the onset of
carbon  ignition  (solid  line), and the  evolution of the center of the
isolated 10 $M_\odot$  star  (dashed  line).  As one can expect from the
previous discussion, the differences between both cases start to show up
at the onset of helium burning at the center, which, as explained above,
is almost coincident with the first mass loss episode.  The evolution of
the primary component of the close binary system leads to higher central
densities  for the same  value  of the  temperature  than  its  isolated
counterpart.  In fact, mass loss has two obvious consequences.  Firstly,
the size of the central helium--burning convective region is smaller for
the case studied  here than in the  evolution  of isolated 10  $M_\odot$
star.  Secondly, the mass of the primary  component of the binary system
is  much  smaller  than  its  isolated   counterpart.  Accordingly,  the
He--exhausted core of the primary is smaller ($\sim 1.05\, M_\odot$) and
with higher central densities than the core resulting from the evolution
of a single 10 $M_\odot$  star ($\sim 1.15\,  M_\odot$) and very similar
to that of a 9 $M_\odot$ isolated star ($\sim 1.04 \, M_\odot$).

The  abundances in the central  regions of the core are also affected by
mass loss in a similar way, and so, the  resulting  composition  is more
similar  to that of the  single  9  $M_\odot$  model.  At the end of the
second mass loss episode the mass of the primary is 1.88 $M_{\odot}$, of
which the innermost 1.05  $M_{\odot}$  corresponds  to a  carbon--oxygen
core  which is  surrounded  by a  helium--rich  envelope.  The  chemical
composition  profiles  at this  moment  are shown in Fig.  8.  As can be
seen in this figure the  abundance  profiles of our model before  carbon
ignition  reveal a higher carbon and neon content, and lower  amounts of
oxygen and magnesium than those of the single 10 $M_\odot$  model.  Also
the helium  burning  shell is narrower in the case studied  here, due to
the fact that the  (helium)  envelope  over this  burning  shell is much
smaller.

%_______________________________________________________________________

\begin{figure}[ht]
\vspace{7.8cm}
\hspace{-2.0cm}
\includegraphics{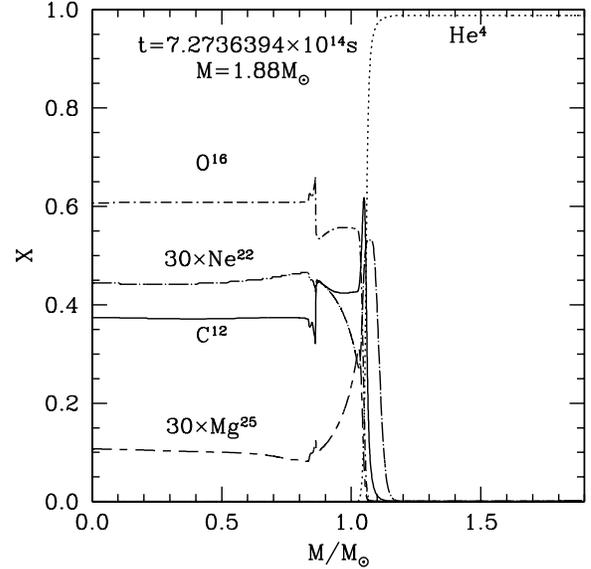}
\caption[]{Abundance  profile of the core of the  primary  at the end of
the second mass loss episode, which is  practically  coincidential  with
the beginning of carbon burning.}
\end{figure}

\section{The carbon burning phase}

The carbon  burning  phase  takes  place  under  conditions  of  partial
degeneracy  and --- as one can expect from the  comparison  between  the
characteristics of the primary star in a close binary system and that of
single  star  models of similar  mass  before  carbon  burning  --- many
similarities  are also found when analysing this phase of the evolution.
Actually, the most prominent  features,  such as the carbon  flashes and
the convective  regions  associated with each one of these (see Fig.  9)
are very  similar for the case  studied  here and those of a 9 $M_\odot$
model star.  These flashes reach about $10^7 - 10^8 \,  L_{\odot}$,  and
last for approximately $10^3$ yr.  However, as it happens in the case of
the evolution of isolated  stars of this mass range,  during most of the
carbon burning phase the surface luminosity  remains almost constant and
close to a value of $\log(L/L_\odot)=4.3$.

\begin{figure}[ht]
\vspace{7.8cm}
\hspace{-2.0cm}
\includegraphics{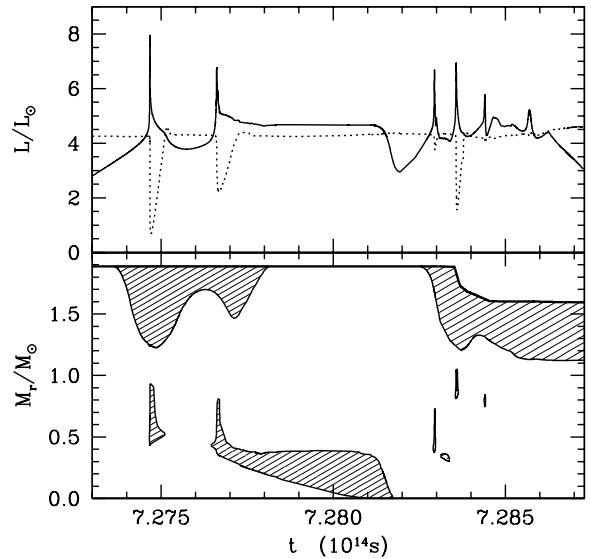}
\caption[]{Evolution    of   the   nuclear   luminosities   during   the
carbon--burning  phase (upper panel), and of the  associated  convective
regions  (lower  panel).  The solid  and the  dotted  lines in the upper
panel correspond,  respectively, to the carbon and helium luminosities.}
\end{figure}

\begin{figure}[ht]
\vspace{9.0cm} 
\hspace{-2.0cm}
\includegraphics{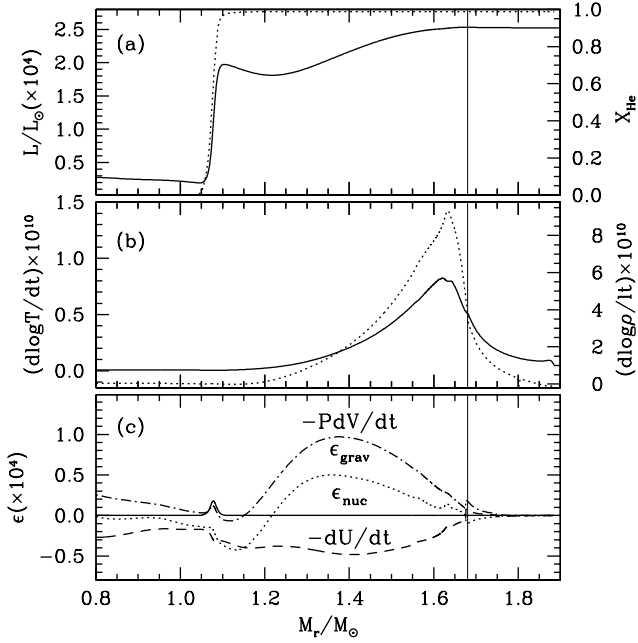}
\caption[]{Relevant  structural and dynamical  quantities for a model at
time $t=7.277763\times  10^{14}$~s, just before the disappearance of the
convective   envelope,   in  a   region   extending   from   below   the
helium--burning shell until the surface of the star.  Panel a:  velocity
(solid line) and helium profile (dotted line).  Panel b:  time variation
of the  temperature  (solid line) and density  (dotted  line).  Panel c:
energy  release  rates  (nuclear and  gravothermal)  along with the time
derivative  of the  internal  energy  and the  work  of  expansion.  The
nuclear  energy  release rate has been divided by $10^2$ in order to fit
into the scale.  The thin solid line marks the position of inner edge of
the convective envelope.}
\end{figure}

The first flash is a prototypic  one.  As in the case of isolated  stars
within this mass range,  carbon is ignited  off--center  due to neutrino
cooling of the central  regions.  The mass coordinate at which carbon is
ignited is $\sim 0.42\,  M_\odot$, very close to the value  obtained for
the  single  9  $M_\odot$  star  ($\sim  0.43\,   M_\odot$).  As  carbon
luminosity  increases, most of the energy generated by nuclear reactions
is used to increase the temperature of the adjacent layers, thus forcing
high temperature  gradients, which ultimately lead to the formation of a
convective zone that allows a more efficient  transport of energy.  This
increase in the  temperature  is followed by an expansion  of the nearby
layers and in particular, of the helium burning  shell, which cools down
and,  hence,  its  luminosity  temporarily  decreases.  Apart  from  the
formation  of the inner  convective  shell, the flash also  affects  the
location of the inner edge of the convective  envelope that moves deeper
into the star when the carbon luminosity  increases and receeds when the
carbon  luminosity  decreases again, very much in the same fashion as in
the isolated 9 $M_\odot$ star.

\begin{figure}[ht]
\vspace{9.0cm}
\hspace{-2.0cm}
\includegraphics{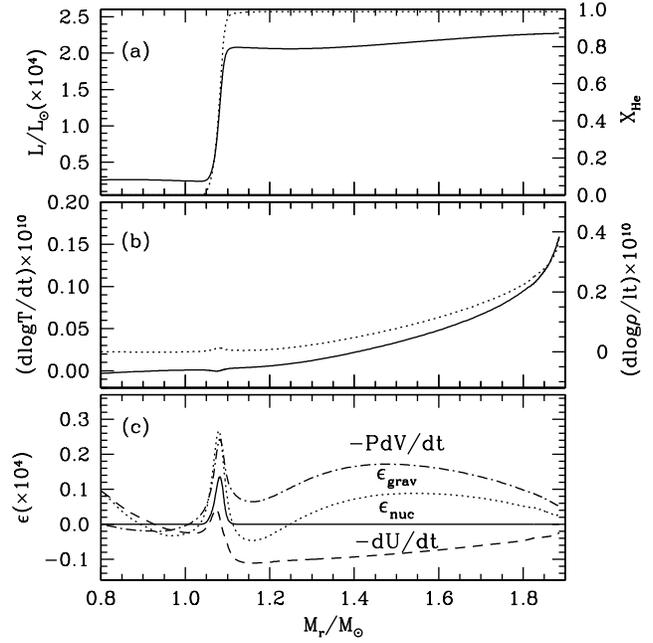}
\caption[]{Same as figure 10, but for time $t=7.278391\times 10^{14}$~s,
just after the convective episode.}
\end{figure}

The  second  flash  occurs  at a  smaller  mass  coordinate  (carbon  is
re--ignited at the point where the  penetration of the inner edge of the
previous  convective  shell was  maximum)  and the  physical  mechanisms
operating  are very  similar  to those of the first  one.  However,  its
maximum strength is considerably  smaller ($\sim 10^7\, L_\odot$ instead
of $\sim 10^8\,  L_\odot$).  After the most violent  phase of this flash
is over, the carbon  luminosity  does not decrease  below that of helium
but, rather, keeps a stationary value,  $\log(L_{\rm  C}/L_\odot) \simeq
4.6$, which is slightly larger than the value of the helium  luminosity,
$\log(L_{\rm   He}/L_\odot)   \simeq   4.3$.  During   this   phase  the
carbon--burning flame propagates inwards at a roughly constant speed ---
see Garc\'\i  a--Berro et al.  (1997) for a detailed  description of the
energy balance  established during this phase --- reaching the center at
$t\simeq 7.282\times 10^{14}$~s.

A very distinctive feature of the case studied here when compared to the
evolution  of  an  isolated  9  $M_\odot$   star  is  that  at  $t\simeq
7.278\times  10^{14}$~s the outer convective envelope  disappears as the
carbon  burning  front  advances  to the  center.  In fact,  the  second
dredge--up in the isolated 9 $M_\odot$  star is caused by the  expansion
and cooling of the layers just below the base of the convective envelope
(Garc\'\i  a--Berro  et al.  1997),  which  allows the  increase  of the
radiative temperature gradient in this region and, so, the inner advance
of  convection  down to $M_{\rm r} \simeq  1.2\,M_{\odot}$.  After that,
the base of the convective envelope remains at an approximately constant
position,  as the energy  supplied by the helium  burning  shell and the
opacity of the nearby  layers  can keep the  temperature  gradient  high
enough.  In the case studied here, the base of the  convective  envelope
reaches  the  same  position  as in the  single  star  due to a  similar
mechanism,  but, unlike the case of the isolated 9 $M_\odot$  star, this
position cannot be maintained,  instead, it receeds and disappears.  The
reason for this  behaviour can be explained  with the help of figures 10
and 11, where we show some relevant structural and dynamical  quantities
for times $t=7.277763\times 10^{14}$~s, just before the disappearance of
the convective  envelope, and  $t=7.278391\times  10^{14}$~s, just after
the convective episode.

Panel {\tt c} of figure 10 shows that  there are three  interesting  and
distinct  regions in the star.  The first of these  regions is below the
helium  discontinuity  and there,  the work of  expansion  fed by carbon
burning is  devoted  basically  to lift  degeneracy  and,  thus, we have
expansion  at almost  constant  temperature  (panel {\tt b}).  On top of
this region we have the helium  burning shell where the energy  supplied
by nuclear  reactions is used to build up the luminosity  profile (panel
{\tt a}).  In the region  between the helium  burning shell and the base
of the  convective  envelope,  the flux is  partially  transformed  into
heating and, at the same time, the whole region is  collapsing.  This in
turn  causes a large  temperature  gradient  at $M_{\rm r} \simeq 1.7 \,
M_\odot$.  As the evolution  continues, the  temperature  in this region
steadily increases.  Thus, the temperature  gradient ultimately flattens
and the inner  edge of the  convective  envelope  consequently  receeds.
Panel {\tt c} of figure 11 shows that the nuclear energy released by the
helium  burning  shell  remains  the same,  but now most of this  energy
merely  flows  outwards  (panel {\tt a}), rather than being  transformed
into work of  expansion.  Instead, the whole region on top of the helium
burning shell is compressed  leading to a  non--homogeneous  increase of
the temperature (panel {\tt b}) which effectively erases the temperature
gradient and forces the  disappearance  of the convective  region.  This
translates  into an increase in the luminosity  for $M_{\rm r} \ga 1.4\,
M_\odot$.

When  carbon--burning  in the central regions is over, a series of small
shell burning episodes occurs (Fig.  9).  Each one of these episodes has
a decreasing  strength and leads to the almost  complete  exhaustion  of
carbon in a core of 1.05  $M_\odot$.  It is  interesting  to note  that,
unlike what happens  with the rest of these mild  flashes,  the first of
these  is not  accompanied  by a  substantial  decrease  in  the  helium
luminosity.  The reason for this  behaviour  is  twofold.  Firstly,  the
duration of this flash is smaller  and, hence, less energy is  released.
Secondly, and most important,  there exists a thick radiative  layer (of
about 0.7  $M_{\odot}$)  between the initial  location of the convective
carbon--burning  shell and the  helium--burning  shell.  Therefore,  the
energy is almost totally  absorbed before  reaching the  helium--burning
shell and, thus, no expansion  nor cooling of the helium  burning  shell
are produced.  During all this phase the core  contracts,  and the outer
layers  of the  star  begin  a new  expansion  as the  outer  convective
envelope  reappears  and its inner edge  advances to the interior of the
star.  However,  the  radius of the star  remains  below the Roche  lobe
radius  and only  exceeds  its value at the end of the  second  of these
flashes.

\begin{figure}[ht]
\vspace{7.8cm}
\hspace{-2.0cm}
\includegraphics{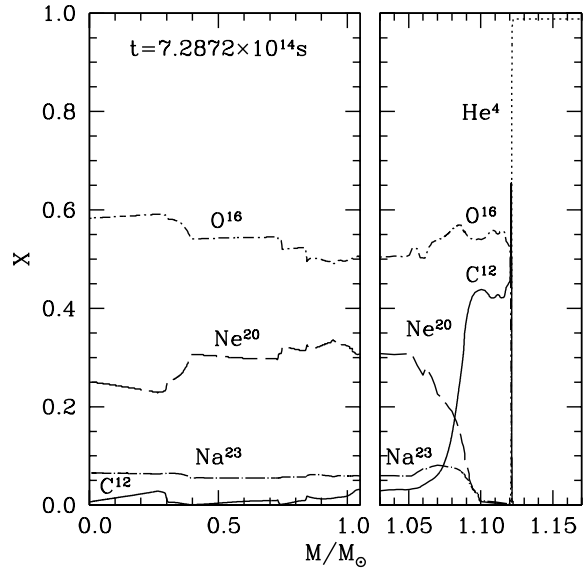}
\caption[]{Final abundances of the primary star after the carbon burning
phase.  Note the existence of a carbon--rich buffer on top of the carbon
exhausted core}
\end{figure}

Since, as  discussed  above,  the  second  mass  loss  episode  was most
probably conservative and during the bulk of the carbon burning phase in
the central regions (say $t \la 7.282 \times  10^{14}$~s) the model star
keeps its radius  below that of the Roche lobe, the  orbital  parameters
remain  unaffected.  As  explained  before,  only at the very end of the
carbon  burning  phase the  flash  activity  is  accompanied  by a rapid
increase of the surface radius of the star beyond the Roche lobe radius.
This, of course, translates into a new mass loss episode for $t\ga 7.283
\times  10^{14}$~s.  This mass loss episode  lasts for a short period of
time, of about $\sim 2 \times  10^{11}$~s and the mass loss rates are of
$\sim  10^{-4}\,  M_\odot\,  {\rm  yr}^{-1}$  at  the  beginning  of the
episode,  becoming much smaller as the  evolution  proceeds.  After this
short phase of Roche lobe overflow, the radius of the primary star turns
out to be very large, but its rate of increase slows down  considerably.
At this evolutionary  stage, the density of the extended  envelope is so
small that a stellar  wind could  also  drive the loss of the  remaining
envelope.  In any case, stellar winds play a significant  role which can
be  comparable  to Roche lobe  overflow.  Thus, we assume that a stellar
wind removes mass from the surface following closely the parametrization
of Nieuwenheuzen  and de Jager (1990) which gives typical rates of $\sim
5  \times  10^{-7}\,  M_\odot  \,  {\rm  yr}^{-1}$  (this  is  the  rate
considered in Fig.  9).  However, since these mass loss rates are poorly
known, we have conducted a series of numerical experiments where we have
changed the mass loss rate from $10^{-5}\,  M_\odot \, {\rm yr}^{-1}$ to
$10^{-7}\,  M_\odot \,{\rm  yr}^{-1}$ and we have found  essentially the
same results for the very final part of the carbon burning phase, except
of course for the mass of the remaining helium envelope.

The  exhaustion of central  carbon is not complete but,  instead, at the
innermost  0.3  $M_{\odot}$  there  remains  remains  a small  amount of
unburnt carbon,  reaching a maximum  abundance of $X_{\rm C} = 0.025$ at
$M_{\rm r}= 0.25\, M_\odot$ (see Fig.  12).  Analogous to what was found
in the series of papers  devoted  to the  evolution  of  isolated  stars
within this mass range, most of the ashes of carbon  burning are sodium,
neon and oxygen, the  abundance  of  magnesium  being very  small.  This
could have important  consequences for the  accretion--induced  collapse
scenario  (Guti\'errez  et al., 1996).  It is also  noteworthy  that the
relatively  thick  carbon--rich  buffer is located just below the helium
envelope,   which  could  make  the  resulting   He--rich   white  dwarf
practically indistinguishable from an observational point of view from a
regular massive carbon--oxygen white dwarf (Weidemann 2000).

%_______________________________________________________________________

\section{Summary and discussion}

We have followed the evolution of a $10\,M_{\odot}$ primary component of
solar metallicity belonging to a close binary system with a secondary of
low or intermediate--low mass, from its main sequence phase until carbon
is exhausted in the core and a  degenerate  remnant is formed.  Our main
goal consisted of extending the previously existing calculations of this
kind of system in order to follow  in full  detail  the  carbon  burning
phase,  which at present  has not yet been well  studied.  We have found
that the primary star undergoes three mass--loss episodes.  The first of
these  episodes  occurs after  hydrogen  exhaustion in the core and most
problably is not conservative, due to the existence of a deep convective
envelope which leads to the formation of a common  envelope.  The second
happens after helium exhaustion, and it is most problably  conservative.
Finally,  the third  one  occurs  when  carbon  burning  in a series  of
succesive shells sets in.  We have analysed the changes in the structure
and  composition  that the primary  star  suffers  while  simultaneously
undergoing carbon burning and mass loss.  The  determination of the mass
loss  rates at this very late  stage of the  evolution,  whether  due to
Roche lobe overflow, or to stellar winds, is still an open question, and
therefore,  we have  explored a broad  range of mass loss rates  ranging
from  $10^{-7}$  to $10^{-5}  M_{\odot}  \, {\rm  yr}^{-1}$  in order to
understand  how the  uncertainties  affect the gross  properties  of the
resulting   degenerate   object.  We  have  found  that  our  incomplete
knowledge of the mass--loss rate does not affect the final result as far
as the  core is  concerned.  In  particular,  carbon  is not  reignited.
Thus, the final  characteristics  of the  remnant  are well  determined.
After the evolution through these different burning stages and under the
influence of a close companion, the initially  $10\,M_{\odot}$ star will
form a $\sim 1.1\,  M_\odot$  degenerate  remnant, with an  oxygen--neon
core of $\sim 0.9\, M_\odot$ surrounded by a carbon--oxygen  rich mantle
of $\sim 0.2\, M_\odot$ and a thin helium envelope.

We have found remarkable differences between our results and those found
by  other  authors  that  have  also  followed  the  evolution  of  a 10
$M_{\odot}$  model star up to the carbon  burning  phase.  For instance,
Nomoto  (1982, 1984)  obtained a final core of larger mass ($\sim  1.3\,
M_\odot$) with significantly higher values for the abundances  Ne$^{20}$
and  Mg$^{24}$,   and  lower  values  for  the  O$^{16}$  and  Na$^{23}$
abundances.  The ultimate reason for this difference is that the nuclear
reaction rates used by Nomoto (Fowler, Caughlan \& Zimmermann, 1975) are
different from ours  (Caughlan \& Fowler,  1988).  In this regard, it is
important  to recall  here that in this  paper we have  chosen  the same
physical inputs as in Ritossa, Garc\'\i a--Berro \& Iben (1996) in order
to remain  consistent with our previous  calculations.  Since then there
have been some new  determinations of the thermonuclear  reaction rates,
particularly the NACRE compilation  (Angulo et al., 1999).  However, for
the most important  reaction  channels  involved in carbon  burning, the
differences  are not  expected  to be  large  and,  thus,  the  chemical
composition  of the core can be considered as relatively  safe,  perhaps
the most important  expected  difference being an even smaller amount of
Mg$^{24}$  (Palacios et al., 2000;  Jos\'e, Coc \& Hernanz,  1999).  The
comparison with the ONe core obtained by  Dom\'{\i}nguez  et al.  (1993)
it is not easy either, since they start from a different initial CO core
prior to carbon  burning.  Moreover,  they  removed  the outer  envelope
during the carbon  burning  phase and, hence, their  treatment  for this
phase differs  substantially  from ours.  This, in particular,  could be
the reason  for the  central  region of  unburnt  carbon  which is quite
apparent in their results,  although poor mass resolution of the central
regions --- see the  discussion in Garc\'\i  a--Berro et al.  (1997) ---
cannot be totally  discarded.  Regarding the average  abundances  of the
ONe core, they obtained, using similar nuclear  reaction  rates, $X({\rm
O}^{16})=0.72$,  $X({\rm  Ne}^{20})=0.25$  and $X({\rm  Mg}^{24})=0.03$.
The high value of the O$^{16}$ abundance is surprising, whereas the rest
of the abundances are similar.  Moreover,  they did not find  Na$^{23}$,
which is an  important  isotope  in our  calculations.  Thus,  all these
differences  could be due to the use of a  simplified  nuclear  network,
much smaller  than ours.  In summary,  the  composition  of our ONe core
differs from the former results, especially for the Mg$^{24}$, Na$^{23}$
and  Ne$^{20}$  nuclei,  which  are the  elements  onto  which  electron
captures  happen  and,   consequently,  may  substantially   affect  the
determination of the explosive ONe ignition  density.  The fact that the
mass of the ONe core found by these authors  ($\sim 0.93\,  M_\odot$) is
quite similar to that obtained here is, however, encouraging.

The definite final result is not necessarily an oxygen--neon white dwarf,
because the  evolution  of the entire  system has not  stopped  yet.  In
fact,  as the  secondary  component  evolves,  it  could  reach  such  a
dimension  that its radius could exceed that of its Roche lobe, and mass
transfer  could take place onto the remnant.  Depending on the evolution
of the orbital  parameters and,  ultimately, on the mass transfer rates,
different possibilities arise:

\begin{enumerate}

\item A cataclysmic variable could be formed, if the accretion rates are
      below  the  critical  rates  for  hydrogen   burning.  At  typical
      accretion rates of about $10^{-9}  M_{\odot}\,{\rm  yr}^{-1}$ onto
      such a white dwarf, the amount of hydrogen--rich material from the
      secondary that must be accreted  before an outburst  occurs ranges
      between  $10^{-4}$ --- see, for instance, Jos\'e \& Hernanz (1998)
      and references therein --- and $10^{-5} M_{\odot}$ (Schwartzman et
      al., 1994).  If we accept the results of Iben and Fujimoto  (1992)
      that significant  hydrogen abundance extends in a mixture with the
      degenerate   material  down  to  a  distance  of  about   $10^{-5}
      M_{\odot}$  under the  surface,  and that  approximately  $2\times
      10^{-5}  M_{\odot}$ are expelled during each outburst, we conclude
      that  between  3\,000 and 4\,000  outburts  (in a total time of at
      least  3--4~Myr)  must occur before the white dwarf is  definitely
      deprived of its $0.035 \, M_{\odot}$ CO rich layer, and so, before
      significant  amounts of Ne$^{20}$  can be detected in the expelled
      material.

\item If  stable  accretion  is  possible,  the  degenerate  object  can
      increase its mass up to the Chandrasekhar mass, finally leading to
      a weak  supernova  explosion,  activating  the  accretion--induced
      collapse   scenario  and  leaving  a  neutron   star.  A  possible
      observational  counterpart  for this outcome  could be the massive
      radio pulsar PSR J0045-7319 (Kaspi et al.  1994), which belongs to
      a binary system in which its companion is an intermediate mass ($M
      \ga 4  M_{\odot}$)  B star.  More  evolved  systems may include an
      intermediate mass binary system consisting of a pulsar and a white
      dwarf.  In  this   sense,   up  to  seven   pulsars   with   these
      characteristics  have recently been discovered (Edwards \& Bailes,
      2001).  In particular, we consider PSR J1756-5322 as a very likely
      counterpart  because the mass of the white dwarf component is $\ga
      0.55 M_{\odot}$.

\item Finally,  if the mass  transfer  rates are so large that the white
      dwarf is not able to accrete all the mass lost of its companion, a
      common  envelope  would form  again, and most of the mass would be
      lost by the  system,  leaving an ONe white  dwarf  belonging  to a
      close binary system.  A possible observational counterpart to this
      outcome is IK Peg  (Smalley et al.,  1996),  which is a  confirmed
      binary system  composed of an unevolved  star plus a massive white
      dwarf  $(M_{\rm  WD} \simeq  1.15\,  M_\odot)$,  possibly  made of
      oxygen and neon.

\end{enumerate}

%_______________________________________________________________________

\begin{acknowledgements}
Part of this work was  supported  by the  Spanish  DGES  project  number
PB98--1183--C03--02,  by the MCYT grant  AYA2000--1785, by the CIRIT and
by   Sun    MicroSystems    under   the   Academic    Equipment    Grant
AEG--7824--990325--SP.  We also wish to thank J.  Jos\'e  for  carefully
reading the manuscript  and to our referee (P.P.  Eggleton) for his very
valuable comments and suggestions.
\end{acknowledgements}

%_______________________________________________________________________

%_______________________________________________________________________

\end{document}